\documentclass[preprint]{aastex62}

\shorttitle{OSSOS: X. How to use a Survey Simulator}
\shortauthors{Lawler et al.}

\begin{document}

\title{OSSOS: X. How to use a Survey Simulator: Statistical Testing of Dynamical Models Against the Real Kuiper Belt}

\correspondingauthor{S.~M. Lawler}
\email{lawler.astro@gmail.com}

\author[0000-0001-5368-386X]{S.~M. Lawler}
\affiliation{NRC-Herzberg Astronomy and Astrophysics, National Research Council of Canada, Victoria, BC, Canada}

\author[0000-0001-7032-5255]{JJ. Kavelaars}
\affiliation{NRC-Herzberg Astronomy and Astrophysics, National Research Council of Canada, Victoria, BC, Canada}
\affiliation{Department of Physics and Astronomy, University of Victoria, Victoria, BC, Canada}

\author[0000-0003-4143-8589]{M. Alexandersen}
\affiliation{Institute of Astronomy and Astrophysics, Academia Sinica, Taipei, Taiwan}

\author[0000-0003-3257-4490]{M.~T. Bannister}
\affiliation{Astrophysics Research Centre, Queen's University Belfast, Belfast, UK}

\author{B. Gladman}
\affiliation{Department of Physics and Astronomy, The University of British Columbia, Vancouver, BC, Canada}

\author[0000-0003-0407-2266]{J.-M. Petit}
\affiliation{Institut UTINAM, UMR 6213 CNRS-Universit\'e de Franche Comt\'e, Besan\c{c}on, France}

\author[0000-0001-7032-5255]{C. Shankman}
\affiliation{Department of Physics and Astronomy, University of Victoria, Victoria, BC, Canada}
\affiliation{NRC-Herzberg Astronomy and Astrophysics, National Research Council of Canada, Victoria, BC, Canada}
\affiliation{current affiliation: City of Toronto, ON, Canada}

\begin{abstract}

All surveys include observational biases, which makes it impossible to directly compare properties of discovered trans-Neptunian Objects (TNOs) with dynamical models.
However, by carefully keeping track of survey pointings on the sky, detection limits, tracking fractions, and rate cuts, the biases from a survey can be modelled in Survey Simulator software.
A Survey Simulator takes an intrinsic orbital model (from, for example, the output of a dynamical Kuiper belt emplacement simulation) and applies the survey biases, so that the biased simulated objects can be directly compared with real discoveries.
This methodology has been used with great success in the Outer Solar System Origins Survey (OSSOS) and its predecessor surveys. 
In this chapter, we give four examples of ways to use the OSSOS Survey Simulator to gain knowledge about the true structure of the Kuiper Belt.
We demonstrate how to statistically compare different dynamical model outputs with real TNO discoveries, how to quantify detection biases within a TNO population, how to measure intrinsic population sizes, and how to use upper limits from non-detections.
We hope this will provide a framework for dynamical modellers to statistically test the validity of their models.

\end{abstract}

\section{Introduction}

The orbital structure, size frequency distribution and total mass of the trans-Neptunian region of the Solar System is an enigmatic puzzle. 
\citet{Fernandez1980} described an expected distribution for this region based on the mechanisms for the delivery of cometary material into the inner Solar System.
Even before the first Kuiper belt object after Pluto was discovered, \citep[1992 QB$_1$;][]{JewittLuu1993,JewittLuu1995}, it was theorized that dynamical effects produced by the mass contained in this region could in principle be detectable \citep{Hamidetal1968}.
The first discoveries made it clear that extracting precise measurements of the orbital and mass distributions from this zone of the Solar System would require careful analysis.

Major puzzles in the Solar System's history can be explored if one has accurate knowledge of the distribution of material in this zone.  
Examples include: the orbital evolution of Neptune \citep[e.g.][]{Malhotra1993}, the large scale re-ordering of the Solar System \citep[e.g.][]{Thommesetal1999,Gomesetal2005}, the process of planetesimal accretion \citep[e.g.][]{Stern1996,DavisFarinella1997}, the production of cometary size objects via collisional processes \citep[e.g.][]{Stern1995} and their delivery into the inner Solar System \citep{Duncanetal1988}, and the stellar environment in which the Sun formed \citep[e.g.][]{BruniniFernandez1996,KobayashiIda2001}.
Our goal as observers is to test these models and their consequences by comparison to the Solar System as we see it today.
Given the sparse nature of the datasets and the challenges of detecting and tracking trans-Neptunian objects (TNOs), a strong statistical framework is required if we are to distinguish between these various models.

The presence of large-scale biases in the detected sample of TNOs has been apparent since the initial discoveries in the Kuiper belt, and multiple approaches have been used to account for these biases.
\cite{JewittLuu1995} use Monte-Carlo comparisons of Kuiper belt models to their detected sample to estimate the total size of the Kuiper belt, taking into account the flux limits of their survey.
Similarly \citet{Irwinetal95} estimate the flux limits of their searches and use these to weight their detections and, combining those with the results reported in \cite{JewittLuu1995}, provide an estimate of the luminosity function of the region. 
\cite{Gladmanetal1998} provide a Bayesian-based analysis of their detected sample, combined with previously published surveys, to further refine the measurement of the luminosity function of the Kuiper belt.
\citet{Trujilloetal2001} determined the size, inclination and radial distributions of the Kuiper belt by weighting the distribution of observed TNOs based on their detectability and the fraction of the orbits that were contained within the survey fields. 
\cite{Bernsteinetal2004} refined the maximum-likelihood approach when they extended the measurement of the size distribution to smaller scales and determined statistically significant evidence of a break in the shape of the Kuiper belt luminosity function, later developed further by the deeper survey of \citet{FraserKavelaars2009}.
A similar approach is taken in \cite{Adams2010} who make estimates of the underlying sampling by inverting the observed distributions.
Other recent results for Kuiper belt subpopulations include \citet{Schwambetal2009}, who use Monte Carlo sampling to estimate detectability of Sedna-like orbits, and \citet{Parker2015}, who uses an approximate Bayesian computation approach to account for unknown observation biases in the Neptune Trojans.
Each of these methods relies on backing out the underlying distributions from a detected sample.

Carefully measuring the true, unbiased structure of the Kuiper belt provides constraints on exactly how Neptune migrated through the Kuiper belt.
Two main models of Neptune's migration have been proposed and modelled extensively.
Pluto's eccentric, resonant orbit was first explained by a smooth migration model for Neptune \citep{Malhotra1993,Malhotra1995}.
Larger scale simulations \citep{HahnMalhotra2005} showed this to be a viable way to capture many TNOs into Neptune's mean-motion resonances.
The so-called ``Nice model'' was proposed as an alternate way to destroy the proto-Kuiper belt and capture many TNOs into resonances \citep{Levisonetal2008}.
In this model, the giant planets undergo a dynamical instability that causes Neptune to be chaotically scattered onto an eccentric orbit that damps to its current near-circular orbit while scattering TNOs and capturing some into its wide resonances \citep{Tsiganisetal2005}.  
Due to the chaotic nature of this model, reproducing simulations is difficult and many variations on the Nice model exist \citep[e.g.][]{Batyginetal2012, Nesvornyetal2013}.
One very recent and promising variation on the Nice model scenario includes the gravitational effects of fairly large ($\sim$Pluto-sized) bodies that cause Neptune's migration to be ``grainy,'' having small discrete jumps as these larger bodies are scattered \citep{NesvornyVokrouhlicky2016}.
More dramatically, even larger planetary-scale objects could have transited and thus perturbed the young Kuiper belt \citep{Petitetal1999,GladmanChan2006,LykawkaMukai2008,SilsbeeTremaine2017}

The level of detail that must be included in Neptune migration scenarios is increasing with the number of discovered TNOs with well-measured orbits; some recent examples of literature comparisons between detailed dynamical models and TNO orbital distributions are summarized here.
\citet{Batyginetal2011}, \citet{DawsonMurrayClay2012}, and \citet{Morbidellietal2014} all use slightly different observational constraints to place limits on the exact eccentricity and migration distance of Neptune's orbit in order to preserve the orbits of cold classical TNOs as observed today. 
\citet{LawlerKozai} test the observed distribution of Kozai Plutinos against the output from a smooth Neptune migration model \citep{HahnMalhotra2005} and a Nice model simulation \citep{Levisonetal2008}, finding that neither model produces sufficiently high inclinations.
\citet{Nesvorny2015a} shows that the timescale of Neptune's migration phase must be fairly slow ($\gtrsim$10~Myr) in order to replicate the observed TNO inclination distribution, and \citet{Nesvorny2015b} shows that including a ``jump'' in Neptune's semimajor axis evolution can create the ``kernel'' observed in the cold classical TNOs \citep[first discussed in][]{Petitetal2011}.
\citet{Pikeetal2017} compare the output of a Nice model simulation \citep{BrasserMorbidelli2013} with scattering and resonant TNOs, finding that the population ratios are consistent with observations except for the 5:1 resonance, which has far more known TNOs than models would suggest \citep{Pikeetal2015}.
Using the observed wide binary TNOs as a constraint on dynamical evolution suggests that this fragile population formed in-situ \citep{ParkerKavelaars2010} or was emplaced gently into the cold classical region \citep{Fraseretal2017}.
Regardless of the model involved, using a Survey Simulator is the most accurate and statistically powerful way to make use of model TNO distributions from dynamical simulations such as these to gain constraints on the dynamical history of the Solar System.

In this chapter, we discuss what it means for a survey to be ``well-characterized'' (Section~\ref{sec:surveys}) and explain the structure and function of a Survey Simulator (Section~\ref{sec:surveysim}).
In Section~\ref{sec:examples}, we then give four explicit examples of how to use a Survey Simulator, with actual dynamical model output and real TNO data.
We hope this chapter provides an outline for others to follow.  

\section{Well-Characterized Surveys} \label{sec:surveys}

A well-characterized survey is one in which the survey field pointings, depths, and tracking fractions at different magnitudes and on-sky rates of motion have been carefully measured.
The largest well-characterized TNO survey to date is the Outer Solar System Origins Survey \citep[OSSOS;][]{Bannisteretal2016}, which was a large program on the Canada-France-Hawaii Telescope (CFHT) carried out over five years, and was specifically planned with the Survey Simulator framework  in mind.
OSSOS builds on the methodology of three previous well-characterized surveys: 
the Canada-France Ecliptic Plane Survey \citep[CFEPS;][]{Jonesetal2006,Kavelaarsetal2008b,Petitetal2011}, 
the CFEPS high-latitude component \citep{Petitetal2017},
and the survey of \citet{Alexandersenetal2016}.
The survey design is discussed extensively in those papers; the OSSOS survey outcomes and parameters in particular are presented in detail in \citet{Bannisteretal2016} and \citet{Bannisteretal2018}.
We here summarize the main points that are important for creating a well-characterized survey.

These well-characterized surveys are all arranged into observing ``blocks:'' many individual camera fields tiled together into a continuous block on the sky (the observing blocks for OSSOS are each $\sim$10--20 square degrees in size).  
The full observing block is covered during each dark run (when the Moon is closest to new) for the 2 months before, 2 months after, and during the dark run closest to opposition for the observing block.
The observing cadence is important for discovering and tracking TNOs, which change position against the background stars on short timescales.
The time separation between imaging successive camera fields inside each observing block must be long enough that significant motion against background stars has occurred for TNOs, but not so long that the TNOs have moved too far to be easily recovered by eye or by software.  
OSSOS used triplets of images for each camera field, taken over the course of two hours.  
The exposure times are chosen as a careful compromise between photometric depth and limiting trailing\footnote{We note that clever algorithms can be used to obtain accurate photometry from trailed sources \citep[e.g.,][]{Fraseretal2016}.} of these moving sources.
The images are searched for moving objects by an automated and robustly tested moving object pipeline \citep[CFEPS and OSSOS used the software pipeline described in][]{Petitetal2004}, and all TNO candidates discovered by the software are visually inspected.
Prior to searching the images, artificial sources (whose flux and image properties closely mimic the real sample) are inserted into the images.  
The detection of these implanted sources marks the fundamental calibration and characterization of the survey block in photometric depth, detection efficiency, and tracking efficiency.
Each observing block has a measured ``filling factor'' that accounts for the gaps between CCD chips.
Detection and tracking efficiencies are measured at different on-sky rates of motion using implanted sources in the survey images, analysed along with the real TNO data.
This process is repeated for each observing block, so each block has known magnitude limits, filling factors, on-sky coverage, and detection and tracking efficiencies at different on-sky rates of motion.
These are then parameterized and become part of the Survey Simulator.

The tracking of the discovered sample provides another opportunity for biases to enter and the process must be closely monitored.  
A survey done in blocks of fields that are repeated $\sim$monthly removes the need to make orbit predictions based on only a few hours of arc from a single night's discovery observations.  
Such short-arc orbit predictions are notoriously imprecise, and dependence on them ensures that assumptions made regarding orbit distribution will find their way into the detected sample as biases.  
For example, a common assumption for short-arc orbits is a circular orbit.  
If follow-up observations based on this circular orbital prediction are attempted with only a small area of sky coverage, then those orbits whose Keplerian elements match the input assumptions will be preferentially recovered, while those that do not will be preferentially lost, resulting in a discovery bias against non-circular orbits.
Correcting for this type of ephemeris bias is impossible.  
Several of the large-sample TNO surveys had short arcs on a high fraction of the detections; this introduces unknown tracking biases into the sample that cannot be reproduced in a Survey Simulator because the systematic reasons for object loss \citep[ephemeris bias,][]{Jonesetal2010} cannot be modeled as random.
Repeatedly observing the same block of fields, perhaps with some adjustment for the bulk motion of orbits, helps ensure that ephemeris bias is kept to a minimum.   
In the OSSOS project we demonstrate the effectiveness of this approach by managing to track essentially \emph{all} TNOs brighter than the flux limits of the discovery sequences \citep[only 2 out of 840 TNOs were not tracked\footnote{The two objects that were not tracked are $d<$15~AU Centaurs whose high on-sky rates of motion caused them to shear off the fields. The possibility of such shearing loss is accounted for in the Survey Simulator.};][]{Bannisteretal2018}.

\citet{Schwambetal2010} is an example of a large-scale TNO survey from outside our collaboration that is well-characterized.  
It has a high tracking fraction and a published pointing history.  
However, it has a comparatively noisy and low-resolution detection efficiency function, thus we do not include it in our Survey Simulator analysis here.
Other large sample size TNO surveys have either unpublished pointings or indications of low tracking fractions leading to unrecoverable ephemeris bias. 

A well-characterized survey will have flux limits in each observing block from measurements of implanted artificial objects, equal sensitivity to a wide range of orbits, and a known spatial coverage on the sky. 
A Survey Simulator can now be configured to precisely mimic the observing process for this survey.

\subsection{The Basics of a Survey Simulator} \label{sec:surveysim}

A Survey Simulator allows models of intrinsic Kuiper Belt distributions to be forward-biased to replicate the biases inherent in a given well-characterized survey.
These forward-biased simulated distributions can then be directly compared with real TNO detections, and a statement can be made about whether or not a given model is statistically consistent with the known TNOs.
One particular strength of this approach is that the effect of non-detection of certain orbits can be included in the analysis.  
Methods that rely on the inversion of orbital distributions are, by their design, not sensitive to a particular survey's blind spots.

Directly comparing a model with a list of detected TNOs (for example, from the Minor Planet Center database), with a multitude of unknown detection biases, can lead to inaccurate and possibly false conclusions.  
Using a Survey Simulator avoids this problem completely, with the only downside being that comparisons can only be made using TNOs from well-characterized surveys.  
Fortunately, the single OSSOS survey contains over 800 TNOs, and the ensemble of well-characterized affiliated surveys contains over 1100 TNOs with extremely precisely measured orbits \citep{Bannisteretal2018}.
This is roughly one third of the total number of known TNOs, and one half of the TNOs with orbits that are well-determined enough to perform dynamical classification.

At its most basic, a Survey Simulator must produce a list of instantaneous on-sky positions, rates of motion, and apparent magnitudes. 
These are computed by assigning absolute magnitudes to simulated objects with a given distribution of orbits.
These apparent magnitudes, positions, and rates of motion are then evaluated to determine the likelihood of detection by the survey, and a simulated detected distribution of objects is produced. The OSSOS Survey Simulator follows this basic model, but takes into account more realities of survey limitations.
It is the result of refinement of this Survey Simulator software through several different well-characterized surveys: initially the CFEPS pre-survey \citep{Jonesetal2006}, then CFEPS \citep{Kavelaarsetal2008a,Petitetal2011}, then \citet{Alexandersenetal2016}, and finally OSSOS \citep{Bannisteretal2016}.
The OSSOS Survey Simulator software and methodology are now robustly tested, and are presented below.

\subsection{The Details of the OSSOS Survey Simulator} \label{sec:surveysim2}

While the methodology presented here is specific to the OSSOS Survey Simulator, by measuring on sky pointings, magnitude limits and tracking fractions, a Survey Simulator can be built for any survey.
The Survey Simulator for the OSSOS ensemble of well-characterized surveys is available as a package\footnote{\url{https://github.com/OSSOS/SurveySimulator}}, and the list of observed characteristics of the TNOs discovered in these surveys is published in \citet{Bannisteretal2018}.
To forward-bias a distribution of objects to allow statistical comparison with the real TNO discoveries in the surveys, the OSSOS Survey Simulator uses the following steps.
The instantaneous on-sky position, rate of motion, and apparent magnitude are computed from an orbit, position, and $H$-magnitude, and can be written to a file that contains the ``drawn'' simulated objects.
This ``drawn'' file then represents the instantaneous intrinsic distribution of simulated objects.
The Survey Simulator evaluates whether each simulated object falls within one of the observing blocks of the survey, and if so, uses the tracking and detection efficiency files for that observing block to calculate whether this object would be detected.
If it is detected, the properties of this object are written to a file containing the simulated detections.
A very small fraction of on-sky motion rates were detected and not tracked in the real survey (Centaurs sometimes shear off the field due to their high rate of motion), which is accounted for in the Survey Simulator. 
The simulated objects are written to the simulated tracked object file with probabilities reflecting this.

The user must supply the Survey Simulator with a routine that generates an object with orbital elements and an absolute $H$ magnitude. 
How these are generated is free for the user to decide.
The user may edit the source software to include generation of orbits and absolute magnitudes within the Survey Simulator, but it is recommended that a separate script is used in conjunction with the unedited Survey Simulator.  
The software package comes with a few examples, and details are provided in the following paragraph showing how we on the OSSOS team have implemented the generation of simulated TNOs.

The orbital elements of an object can be determined in a variety of ways. 
The Survey Simulator can choose an orbit and a random position within that orbit, either from a list of orbits (as would be produced by a dynamical model) or from a parametric distribution set by the user.
Orbits from a list can also be easily ``smeared,'' that is, variation is allowed within a fraction of the model orbital elements, in order to smooth a distribution or produce additional similar orbits (however, one must be careful that the distribution is dominated by the original list of orbits, and not by the specifics of the smearing procedure). 
To determine the likely observed magnitude of the source, an absolute $H$ magnitude is assigned to this simulated object, chosen so as to replicate an $H$-distribution set by the user.
Tools and examples are provided to set the $H$-distribution as a single exponential distribution, to include a knee to a different slope at a given $H$ magnitude \citep[see, for example,][]{Fraseretal2014}, or to include a divot in the $H$-distribution \citep[as in][]{Shankmanetal2016,Lawleretal2018}
The $H$-distribution parameter file also sets the maximum and minimum $H$ values that will be simulated.  
The smallest $H$-magnitude (i.e.\ largest diameter) is not as important because it is set by the distribution itself.
But the largest $H$-magnitude (i.e.\ smallest diameter) is important to match to the population being simulated.
If the maximum $H$-magnitude is smaller than that of the largest magnitude TNO in the observational sample one wants to simulate, then the distribution won't be sensitive to this large $H$-magnitude tail.
If the maximum $H$ value is much larger than that of the largest magnitude TNO in the sample, it is simply a waste of computational resources, since the simulation will include many objects that were too faint to be detected in the survey.
This does, however, expose a strength of the Survey Simulator, as we can learn from these faint detections that a hidden reservoir of small objects might exist. 

The process of drawing simulated objects and determining if they would have been detected by the given surveys is repeated until the desired number of simulated tracked or detected objects is produced by the Survey Simulator.
The desired number of simulated detected objects may be the same as the number of real detected TNOs in a survey in order to measure an intrinsic population size (as demonstrated in Section~\ref{sec:centaurs}), or an upper limit on a non-detection of a particular subpopulation (Section~\ref{sec:nondet}), or it may be a large number in order to test the rejectability of an underlying theoretical distribution (as demonstrated in Sections~\ref{sec:P9}) or to quantify survey biases in a given subpopulation (Section~\ref{sec:res}).

\section{Examples of Survey Simulator Applications} \label{sec:examples}

Here we present four examples of different ways to use the Survey Simulator to gain statistically valuable information about TNO populations.
In Section~\ref{sec:P9}, we demonstrate how to use the Survey Simulator to forward-bias the output of dynamical simulations and then statistically compare the biased simulation with a distribution of real TNOs.
In Section~\ref{sec:res}, we use the Survey Simulator to build a parametric intrinsic distribution and then bias this distribution by our surveys, examining survey biases for a particular TNO subpopulation in detail.
The Survey Simulator can also be used to measure the size of the intrinsic population required to produce a given number of detections in a survey; this is demonstrated in Section~\ref{sec:centaurs}.
And finally, in Section~\ref{sec:nondet}, we demonstrate a handy aspect of the Survey Simulator: using non-detections from a survey to set statistical upper limits on TNO subpopulations.
We hope that these examples will prove useful for dynamical modellers who want a statistically powerful way to test their models.

\subsection{Testing the Output of a Dynamical Model: The Outer Solar System with a Distant Giant Planet or Rogue Planet} \label{sec:P9}

This example expands on the analysis of \citet{Lawleretal2017}, which presents the results of three different dynamical emplacement simulations of the distant Kuiper Belt.  
Here we analyze these three dynamical simulations and also an additional emplacement model that includes a ``rogue planet'' that is ejected early in the Solar System's history \citep{GladmanChan2006}. The outputs from these four dynamical TNO emplacement simulations are run through the OSSOS Survey Simulator and compared with the high pericenter TNOs discovered by OSSOS.  The four emplacement simulation scenarios analyzed here include the following conditions:
\begin{enumerate}
\item The four giant planets and including the effects of Galactic tides and stellar flybys \citep[simulation from][]{Kaibetal2011}
\item The four giant planets with an additional 10 Earth mass planet having $a=250$~AU and $e=0$ \citep[based on a theory proposed in][]{TrujilloSheppard2014}, also including the effects of Galactic tides and stellar flybys \citep[simulation from][]{Lawleretal2017}.
\item The four giant planets with an additional 10 Earth mass planet having $a=500$~AU and $e=0.5$ \citep[based on a theory proposed in][]{BatyginBrown2016}, also including the effects of Galactic tides and stellar flybys \citep[simulation from][]{Lawleretal2017}.
\item The four giant planets and an additional 2 Earth mass rogue planet that started with $a=35$~AU and $q=30$~AU, which was ejected after $\sim$200~Myr \citep[simulation from][]{GladmanChan2006}.
\end{enumerate} 

The papers which have recently proposed the presence of a distant undiscovered massive planet \citep[popularly referred to as ``Planet 9'';][]{TrujilloSheppard2014,BatyginBrown2016} rely on  published detections of large semimajor axis ($a$), high pericenter distance ($q$) TNOs, which have extreme biases against detection in flux-limited surveys\footnote{\citet{Brown2017} does attempt a numerical simulation to calculate likelihood of detection for different orbital parameters in the high-$q$ population, but this is backing out the underlying distribution in the underlying sample and still relies on assumptions about the unknown sensitivity and completeness of the surveys.}. 
These large-$a$, high-$q$ TNOs are drawn from different surveys, with unpublished and thus unknown observing biases. 
The Minor Planet Center (MPC) Database, which provides a repository for TNO detections, does not include information on the pointings or biases of the surveys that detected these TNOs.
Sample selection caused by survey depth and sky coverage can be non-intuitive, caused by weather, galactic plane avoidance, and telescope allocation pressure, among other possibilities \citep{SheppardTrujillo2016,Shankmanetal2017}.
Thus the assumption can \emph{not} be made that a collection of TNOs from different surveys will be bias-free.  
Further, the detected sample of objects in the MPC Database provides no insight into the parts of the sky where a survey may have looked and found nothing, nor to strong variation in flux sensitivity that occurs due to small variations in image quality.

\begin{figure}[h!]
\begin{center}
\includegraphics[scale=0.5]{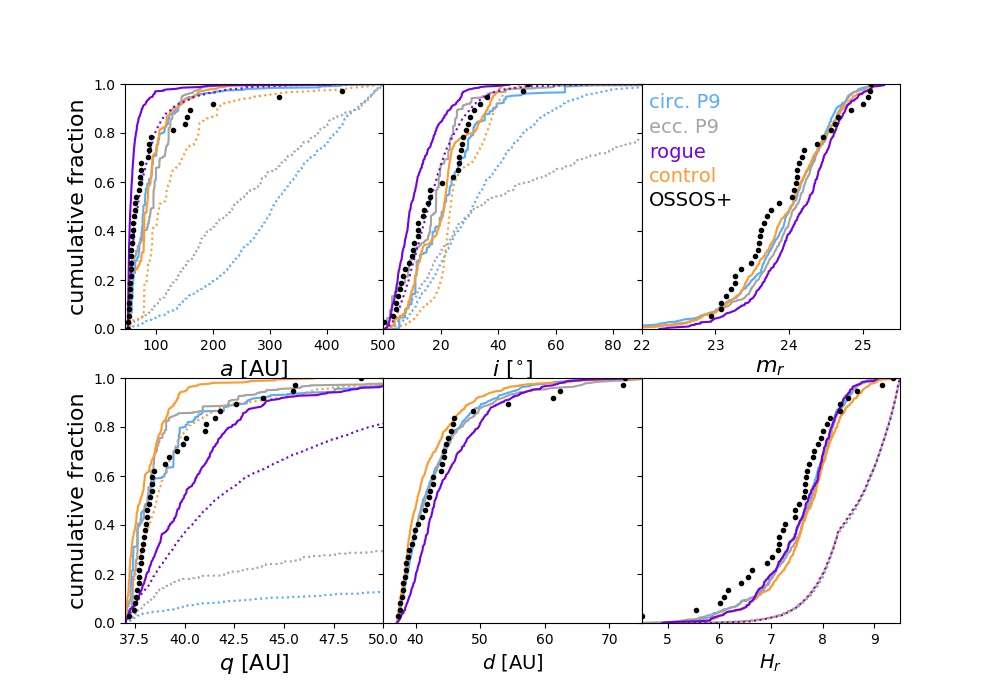}
\end{center}
\caption{Cumulative distributions of TNOs in six different parameters: semimajor axis $a$, inclination $i$, apparent $r$-band magnitude $m_r$, pericenter distance $q$, distance at detection $d$, and absolute magnitude in $r$-band $H_r$.
The result of four different emplacement models are presented: the known Solar System (orange), the Solar System plus a circular orbit Planet~9 (blue), or plus an eccentric orbit Planet~9 (gray), and a ``rogue'' planet simulation (purple).
The intrinsic model distributions are shown with dotted lines, and the resulting simulated detections in solid lines.
Black circles show non-resonant TNOs discovered by the OSSOS ensemble having $q>37$~AU and $50<a<500$~AU.
}\label{fig:cumuplot}
\end{figure}

Here we demonstrate how to use the results of well-characterized surveys to compare real TNO detections to the output of a dynamical model, comprising a list of orbits.
We visualize this output with a set of cumulative distributions.
Figure~\ref{fig:cumuplot} shows cumulative distributions in six different observational parameters: semimajor axis $a$, inclination $i$, apparent magnitude in $r$-band $m_r$, pericenter distance $q$, distance at detection $d$, and absolute magnitude in $r$-band $H_r$.
The outputs from the four emplacement models are shown by different colours in the plot, with the intrinsic distributions shown as dotted lines, and the forward-biased simulated detection distributions as solid lines.   
These intrinsic distributions have all been cut to only include the pericenter and semimajor axis range predicted by \citet{BatyginBrown2016} to be most strongly affected by a distant giant planet: $q>37$~AU and $50<a<500$~AU.

We cannot directly compare the output from these dynamical simulations covering such a huge range of $a$ and $q$ to real TNOs because the observing biases present in the detected TNO distributions are severe.  
Using a survey simulator combined with a carefully characterized survey, however, allows us to impose the observing biases onto the simulated sample and determine what the detected simulated sample would have been.
We use the OSSOS Survey Simulator and the $q>37$~AU, $50<a<500$~AU TNOs discovered by the OSSOS ensemble in this example (Figure~\ref{fig:cumuplot}).

The solid lines in Figure~\ref{fig:cumuplot} show the forward-biased simulated detections from the Survey Simulator, and the black points show the real TNOs detected in the OSSOS ensemble with the same $a$ and $q$ cut.
These can now be directly compared, as the same biases have been applied.
The effect of observing biases varies widely among the six parameters plotted, and can be seen by comparing the intrinsic simulated distributions (dotted lines) with the corresponding biased simulated distributions (solid lines) in Figure~\ref{fig:cumuplot}.
Unsurprisingly, these surveys are biased toward detecting the smallest-$a$, lower inclination, and lowest-$q$ TNOs.  

These models have not been explicitly constructed in an attempt to produce the orbital and magnitude distribution of the elements in the detection range, so the following exercise is pedagogic rather than attempting to be diagnostic.
All these models statistically fail dramatically in producing several of the distributions shown here (discussed in detail later in this section), but how they fail allows one to understand what changes to the model may be required.
All the models generically produce a slight misbalance in both the absolute (lower right panel of Figure~\ref{fig:cumuplot}) and apparent magnitude distributions (upper right panel of Figure~\ref{fig:cumuplot}), with too few bright objects being present in the predicted sample, when using the input $H_r$-magnitude distribution from \citet{Lawleretal2018}.  
Changes in the $H$-magnitude exponents or break points that are comparable to their current uncertainties will be sufficient to produce a greatly improved match, so this comparison is less interesting than for the orbital distributions.

The orbital inclination distribution (upper center panel of Figure~\ref{fig:cumuplot}) of the detected sample (black dots) is roughly uniform up to about $i=18^{\circ}$, it has few members from $18-25^{\circ}$, and then has the final one third of the sample distributed up to about $40^{\circ}$.
This observed cutoff near $40^{\circ}$ is strongly affected by survey biases, as evidenced by the dramatic elimination of this large fraction of the intrinsic model population that is present in the two Planet~9 simulations (compare the distribution of dotted and solid blue and gray lines).
In contrast, the relative dearth in the $18-25^{\circ}$ range of the observed TNOs is {\it not} caused
by the survey biases: none of the biased models show this effect.
The rogue planet model's intrinsic distribution (dotted purple line) is a relatively good match to the detections up to about $i=15^{\circ}$; when biased by the Survey Simulator (solid purple line) this model predicts far too cold a distribution.  
The rogue model used here was from a simulation with a nearly-coplanar initial rogue planet; simulations with an initially inclined extra planet will give higher inclinations for the Kuiper Belt objects and would thus be required in this scenario.
The two Planet~9 scenarios shown give better (but still rejectable) comparisons to the detections.

The comparison of the models in the semimajor axis distribution gives very clear trends (upper left panel of Figure~\ref{fig:cumuplot}).
The rogue planet model (solid purple line) used seriously underpredicts the fraction of large-$a$ TNOs but does produce the abundant $a<70$~AU objects in the detected sample (black dots).
This is caused by the rogue spending little time in the enormous volume beyond 100~AU and thus being unable to significanly lift perihelia for those semimajor axes.
In contrast, the control (solid orange line) and Planet~9 models (solid blue and gray lines) greatly underpredict the $a<100$~AU fraction because the distant planet has very little dynamical effect on these relatively tightly-bound orbits; these models predict roughly the correct fraction, $\sim$20\% of objects having $a>100$~AU, but the distribution of larger-$a$ TNOs is more extended in reality than these models predict.
Lastly, the $q>37$~AU perihelion distribution (lower left panel of Figure~\ref{fig:cumuplot}) of the rogue model (solid purple line) is more extended than the real objects, due to this particular rogue efficiently raising the perihelia of the $a=50-100$~AU populations by a few AU, to a broad distribution. 
The control model's $q$ distribution (solid orange line) is far too concentrated, while the Planet~9 simulations (solid blue and gray lines) qualitatively provide the match the real TNOs in this distribution.

In order to determine whether any of these biased distributions provide a statistically acceptable match to the real TNOs, we use a bootstrapped Anderson-Darling statistic\footnote{The Anderson-Darling test is similar to the better-known Kolmogorov-Smirnov test, but with higher sensitivity to the tails of the distributions being compared.} \citep{AndersonDarling}, calculated for each of the six distributions.
This technique is described in detail in previous literature \citep[e.g.][]{KavelaarsL3,Petitetal2011,Bannisteretal2016,Shankmanetal2016}, and we summarize below.
An Anderson-Darling statistic is calculated for the simulated biased distributions compared with the real TNOs; this statistic is summed over all distributions being tested.  
The Anderson-Darling statistic is then bootstrapped by drawing a handful of random simulated objects from each simulated distribution, calculating the resulting AD statistics, summing over all distributions, then comparing this to how often the summed AD statistic for the real TNOs occurs \citep[following][]{Parker2015, Alexandersenetal2016}.
If a summed AD statistic as large as the summed AD statistic for the real TNOs occurs in $<5$\% of randomly drawn samples, we conclude this simulated distribution is inconsistent with observations and we can reject it at the $95\%$ confidence level. 

When we calculate the bootstrapped AD statistics for each of the simulated distributions as compared with the real TNOs in Figure~\ref{fig:cumuplot}, we find that \emph{all four} of the tested simulations are inconsistent with the data and we reject all of them at $>99\%$ confidence level.
This may be surprising to those unaccustomed to these comparisons.  
Matches between data and models that are not statistically rejectable have almost no noticeable differences between the data and the biased model in all parameters than are compared (see, e.g., Figure~3 in \citealt{Gladmanetal2012}).

We note that none of the four dynamical emplacement models analysed here include the effects of Neptune's migration, which is well-known to have an important influence on the structure of the distant Kuiper belt.
Recent detailed migration simulations \citep{KaibSheppard2016,Nesvornyetal2016,Pikeetal2017,PikeLawler2017} have shown that temporary resonance capture and Kozai cycling within resonances during Neptune's orbital evolution has important effects on the overall distribution of distant TNOs, particularly in raising pericenters and semimajor axes.
Incorporating the effects of Neptune's migration may produce a better fit between the real TNOs and the models shown in Figure~\ref{fig:cumuplot}.

We reiterate that the point of this section has been to provide a walk-through of how to compare the output of a dynamical model to real TNO detections in a statistically powerful way.  
The preceding discussion of the shortcomings of the specific dynamical models presented here highlights that a holistic approach to dynamical simulations of Kuiper belt emplacement is necessary.
For example, \citet{Nesvorny2015a} uses the CFEPS Survey Simulator and CFEPS-discovered TNOs to constrain a Neptune migration model presented in that work, and   
\citet{Shankmanetal2016} uses the OSSOS Survey Simulator and TNOs to improve the dynamical emplacement model of \citealt{Kaibetal2011}.
We hope that use of a Survey Simulator will become standard practice for testing dynamical emplacement models in the future.

\subsection{Using a Parametric Population Distribution: Biases in Detection of Resonant TNOs} \label{sec:res}

In this section we show how to use the Survey Simulator with a resonant TNO population in order to demonstrate two important and useful points: how to build a simulated population from a parametric distribution, and showing how the Survey Simulator handles longitude biases in non-uniform populations.
For this demonstration, we build a toy model of the 2:1 mean motion resonance with Neptune, using a parametric distribution built within the Survey Simulator software (though we note that this parametric distribution could just as easily be created with a separate script and later utilized by the unedited Survey Simulator).
The parametric distribution here is roughly based on that used in \citet{Gladmanetal2012}, but greatly simplified and not attempting to match the real 2:1 TNO distributions.
The Survey Simulator chooses from this parametric distribution with $a$ within $\pm0.5$~AU of the resonance center (47.8~AU), $q=30$~AU, $i=0^{\circ}$,\footnote{The inclinations are actually set to a very small value close to zero to avoid ambiguities in the other orbital angles.} and libration amplitudes $0-10^{\circ}$. 
There are three resonant islands in the 2:1 resonance, and the libration center $\langle\phi_{21}\rangle$ is chosen to populate these three islands equally in this toy model. 
For the purposes of our toy model, we define these angles as: leading asymmetric $\langle\phi_{21}\rangle=80^{\circ}$, symmetric $\langle\phi_{21}\rangle=180^{\circ}$, and trailing asymmetric $\langle\phi_{21}\rangle=280^{\circ}$.
The resonant angle $\phi_{21}$ in the snapshot is chosen sinusoidally within the libration amplitude.  
The ascending node $\Omega$ and mean anomaly $\mathcal{M}$ are chosen randomly, then the argument of pericenter $\omega$ is chosen to satisfy the resonant condition $\phi_{21}=2\lambda_{\rm TNO}-\lambda_{\rm N}-\varpi_{\rm TNO}$, where mean longitude $\lambda=\Omega+\omega+\mathcal{M}$, longitude of pericenter $\varpi=\Omega+\omega$, and the subscripts TNO and N denote the orbital elements of the TNO and Neptune, respectively.
The last step is to assign an $H$ magnitude, in this case from a literature TNO $H$-distribution \citep{Lawleretal2018}.
As each simulated TNO is drawn from this distribution, its magnitude (resulting from its instantaneous distance and $H$ magnitude), on-sky position, and rate of on-sky motion are evaluated by the Survey Simulator to ascertain whether or not this TNO would have been detected by the survey for which configurations are provided to the simulator.

\begin{figure}[h!]
\begin{center}
\includegraphics[scale=0.7]{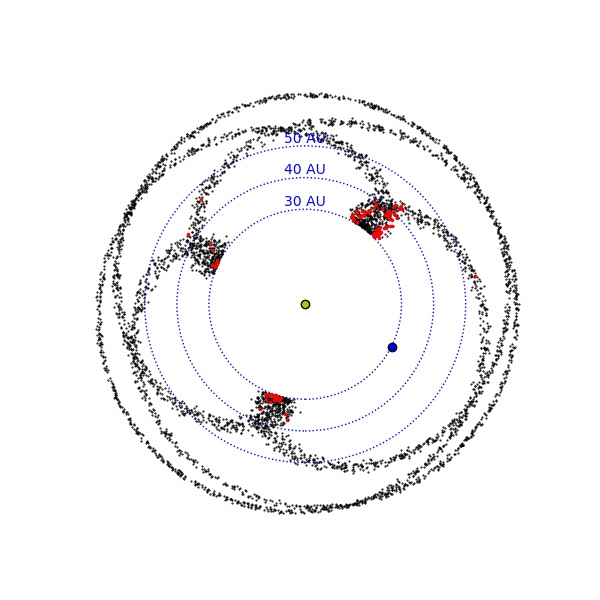}
\end{center}
\caption{Black points show the positions of TNOs in a snapshot from this parametric toy model of TNOs in the 2:1 mean-motion resonance.
The position of Neptune is shown by a blue circle, and dotted circles show distances from the Sun.
Red points show simulated detections after running this model through the Survey Simulator. 
One third of the TNOs are in each libration island in the intrinsic model, but 2/3 of the detections are in the leading asymmetric island (having pericenters in the upper right quadrant of this plot).
This bias is simply due to the longitudinal direction of pointings within the OSSOS ensemble and pericenter locations in this toy model of the 2:1 resonance.
}\label{fig:toy21}
\end{figure}

Figure~\ref{fig:toy21} shows the results of this simulation.  
Due to the (unrealistically) low libration amplitudes in this toy model simulation, the three resonant islands show up as discrete sets of orbits, with pericenters in three clusters: symmetric librators are opposite Neptune, and leading and trailing are in the upper right and lower left of the plot, respectively\footnote{We note that in reality the symmetric island generally has very large libration amplitudes and the pericenter positions of symmetric librators overlap with each of the asymmetric islands \citep[e.g.][]{Volketal2016}, thus diagnosing the membership of each island is not as simple as this toy model makes it appear.}.
While the intrinsic distributions in this model are evenly distributed among the three islands by design (33\% in each), the detected TNOs are not.
This is because TNOs on eccentric orbits are most likely to be detected close to pericenter, and in this toy model, the pericenters are highly clustered around three distinct positions on the sky.\footnote{While this is obviously an exaggerated example, pericenters in \emph{all} of the resonant TNO populations are most likely to occur at certain sky positions, and thus this is an important consideration when discussing survey biases \citep[see][for detailed discussions and analysis of these effects for resonant populations]{Gladmanetal2012,LawlerKozai,Volketal2016}.}

The red points in Figure~\ref{fig:toy21} show the simulated TNOs that are detected by the OSSOS Survey Simulator.  
Just because of the positions of survey pointings on the sky, the detections are highly biased toward the leading island, which has over 2/3 of the detections, with fewer detections in the trailing island, and only 10\% of all the detections in the symmetric island.  
Without knowing the on-sky detection biases (as is the case for TNOs pulled from the MPC database), one could easily (but erroneously) conclude that the leading island is much more populated than the trailing island, when in fact the intrinsic populations are equal.

The relative fraction of $n$:1 TNOs that inhabit different resonant islands has been discussed in the literature as a diagnostic of Neptune's past migration history \citep{ChiangJordan2002,MurrayClayChiang2005,PikeLawler2017}.
Past observational studies have relied on $n$:1 resonators from the MPC database with unknown biases, or the handful of $n$:1 resonators detected in well-characterized surveys to gain weak (but statistically tested) constraints on the relative populations.  
Upcoming detailed analysis using the full OSSOS survey and making use of the Survey Simulator will provide stronger constraints on this fraction, without worries about unknown observational biases (Chen, Y.-T., private communication).

\subsection{Determining Intrinsic Population Size: The Centaurs} \label{sec:centaurs}

The Survey Simulator can easily be used to determine intrinsic population sizes.  
As described in Section~\ref{sec:surveysim}, the Survey Simulator keeps track of the number of ``drawn'' simulated objects that are needed for the requested number of simulated tracked objects.
By asking the Survey Simulator to produce the same number of tracked simulated objects as were tracked in a given survey, the number of drawn simulated objects is a realization of the intrinsic population required for the survey to have detected the actual number of TNOs that were found by the survey.
By repeating this many times, with different random number seeds, different orbits and instantaneous positions are chosen from the model and slightly different numbers of simulated drawn objects are required each time.
This allows us to measure the range of intrinsic population sizes needed to produce a given number of tracked, detected TNOs in a survey.

\begin{figure}[h!]
\begin{center}
\includegraphics[scale=0.6]{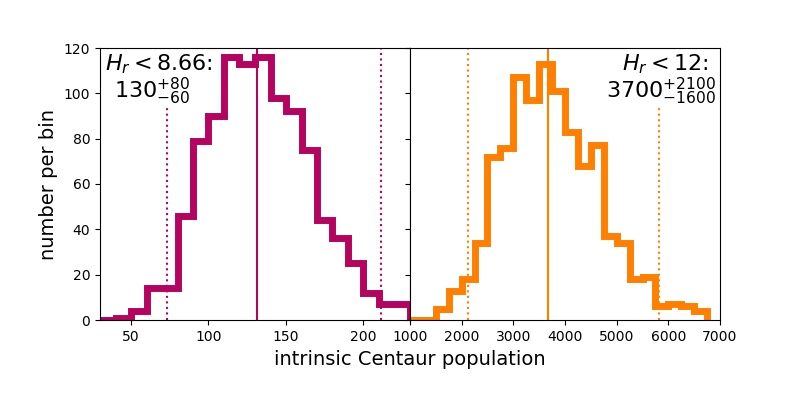}
\end{center}
\caption{The range of intrinsic Centaur population sizes required for the Survey Simulator to produce the same number of Centaur detections (17) as were discovered by the OSSOS ensemble, $H_r<8.66$ intrinsic populations shown in red (left panel) and $H_r<12$ shown in orange (right panel).
The solid lines highlight the median population, and dotted lines show the 95\% upper and lower confidence limits on the intrinsic Centaur population for each $H$-magnitude limit.
Using the OSSOS ensemble, we measure an intrinsic populations of $130^{+80}_{-60}$ $H_r<8.66$ Centaurs and $3700^{+2100}_{-1600}$ $H_r<12$ Centaurs with 95\% confidence.
}\label{fig:cenpop}
\end{figure}

Here we measure the intrinsic population required to produce the 17 Centaurs that are detected in the OSSOS ensemble.  
Once the parameters of the orbital distribution and the $H$-distribution have been pinned down using the AD statistical analysis outlined in Section~\ref{sec:P9} \citep[this is done in detail for the Centaurs in][]{Lawleretal2018}, we run the Survey Simulator until it produces 17 tracked Centaurs from the $a<30$~AU portion of the \citet{Kaibetal2011} scattering TNO model, and record the number of simulated drawn objects required.
As the OSSOS ensemble discovered Centaurs down to $H_r\simeq14$, the Survey Simulator is run to this $H$-magnitude limit.
We repeat this 1000 times to find the range of intrinsic population sizes that can produce 17 simulated tracked objects.

Using the properties of simulated objects in the drawn file, we measure the intrinsic population size to $H_r<12$ (right panel, Figure~\ref{fig:cenpop}).  
The median intrinsic population required for 17 detections is 3700, with 97.5\% of population estimates falling above 2100, and 97.5\% falling below 5800 (these two values bracket 95\% of the population estimates).
The result is a statistically tested 95\% confidence limit on the intrinsic Centaur population of $3700^{+2100}_{-1600}$ for $H_r<12$.

To ease comparison with other statistically produced intrinsic TNO population estimates \citep[e.g.][]{Petitetal2011,Gladmanetal2012}, we also measure a population estimate for $H_r<8.66$, which corresponds to $D\gtrsim100$~km\footnote{The approximate $H_r$ magnitude that corresponds to $D$ of 100~km is calculated assuming an albedo of 0.04 and using an average plutino colour $g-r=0.5$ \citep{Alexandersenetal2016}.} (left panel, Figure~\ref{fig:cenpop}). 
Even though none of the detected Centaurs in OSSOS had $H_r$ magnitudes this small, this estimate is valid because it is calculated from our Survey Simulator-based population estimate and a measured $H$-magnitude distribution \citep[from][]{Lawleretal2018}.
For $H_r<8.66$, the statistically tested 95\% confidence limit on the intrinsic Centaur population is $130^{+80}_{-60}$.

\subsection{Constraints from Non-Detections: Testing a Theoretical Distant Population} \label{sec:nondet}

Here we show a perhaps unintuitive aspect of the Survey Simulator: non-detections can be just as powerful as detections for constraining Kuiper Belt populations.
Non-detections can only be used if the full pointing list from a survey is published along with the detected TNOs.
An examination of the orbital distribution of TNOs in the MPC database makes it clear that there is a sharp dropoff in the density of TNO detections at $a\gtrsim50$~AU and $q\gtrsim40$~AU \citep[Figure~2 in][demonstrates this beautifully]{SheppardTrujillo2016}.
Is this the result of observation bias or a real dropoff?
Without carefully accounting for survey biases \citep{Allenetal2001,Allenetal2002}, by using the MPC database alone, there is no way to know.

\begin{figure}[h!]
\begin{center}
\includegraphics[width=6.5in]{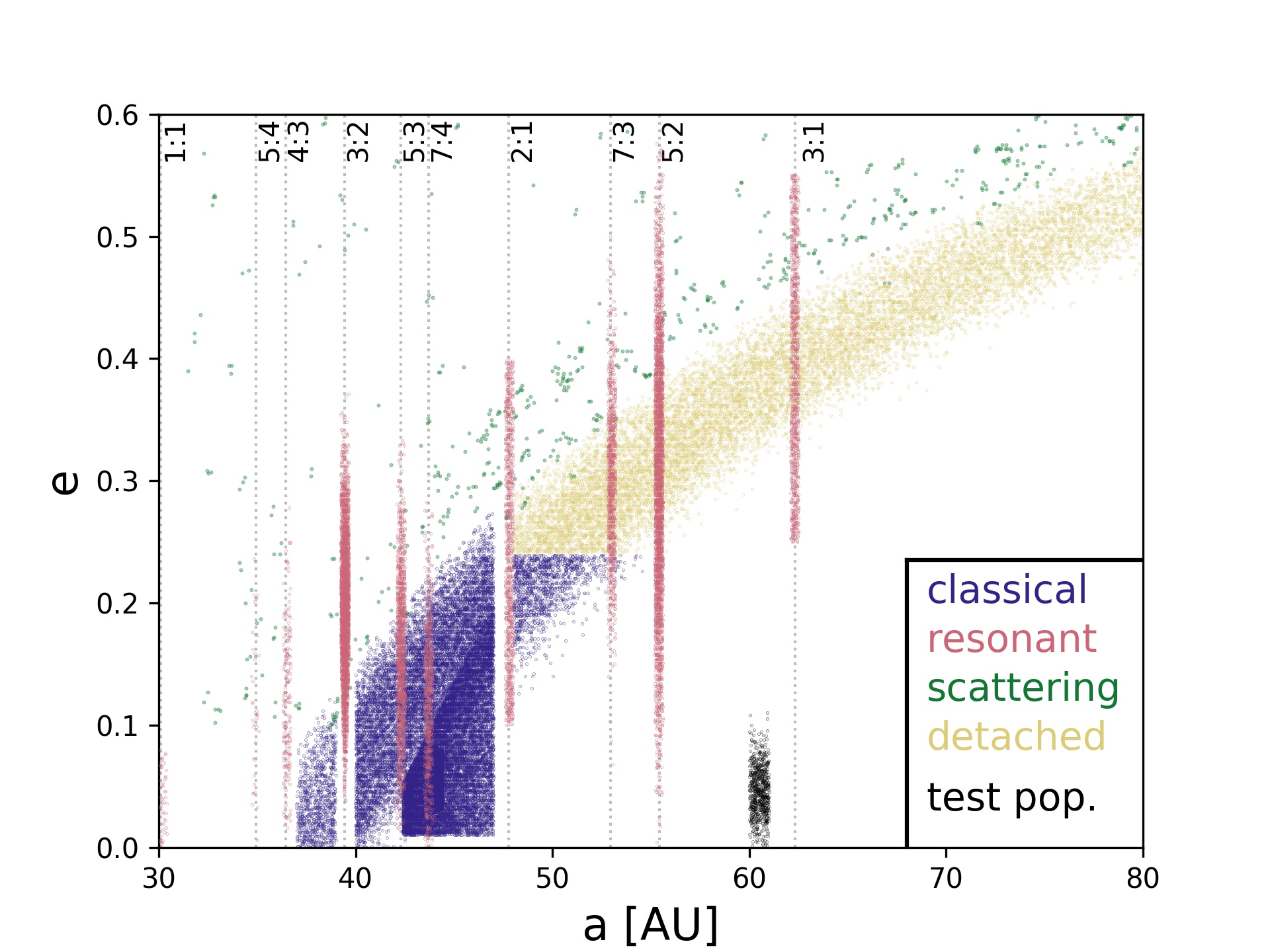}
\end{center}
\caption{
Coloured points show the relative populations and semimajor axis-eccentricity distributions from the CFEPS L7 model of the Kuiper belt, where absolute population estimates have been produced for each subpopulation in the well-characterized CFEPS survey \citep{Petitetal2011,Gladmanetal2012}, populations are scaled to $H_g<8.5$ and colour-coded by dynamical class.
Resonances included in this model (those with $>$1 TNO detected by CFEPS) are labelled.
A low-$e$ artificial test population has been injected at 60~AU (black points); the number of points shows the upper limit on this population determined by zero detections in OSSOS, extrapolated to match the CFEPS model ($H_r<8$).
}\label{fig:ring}
\end{figure}

Randomly drawing zero objects when you expect three has a probability of 5\%, assuming Poisson statistics; thus, the simulated population required to produce three detections is the 95\% confidence upper limit for a population that produced zero detections. 
As our example for non-detection upper limits, we create an artificial population in the distant, low-eccentricity Kuiper belt, where OSSOS has zero detections.
Figure~\ref{fig:ring} shows the CFEPS L7 debiased Kuiper belt\footnote{Model available at \url{www.cfeps.net}} \citep{Petitetal2011,Gladmanetal2012}, where this low-eccentricity artificial population has been inserted at 60~AU (small black points) to demonstrate the power of non-detections, even in parameter space with low sensitivity (i.e., low-$e$ TNOs at 60~AU will never come very close to the Sun, thus will always remain faint and difficult to detect).
We run this artificial population through the Survey Simulator until we have three detections in order to measure a 95\% confidence upper limit on this population.
At this large distance, we are only sensitive to relatively large TNOs ($H_r\lesssim7$, corresponding to $D\gtrsim180$~km for an albedo of 0.04). 
For there to be an expectation to detect three objects in the survey fields, the Survey Simulator tells us that there would need to exist a population of $\sim$90 such objects, which is our upper limit on this population.  
Extrapolating this population to match the $H$ magnitude limits of the CFEPS L7 subpopulations plotted in Figure~\ref{fig:ring}, the total number of which are scaled appropriately for $H_g<8.5$, gives $\sim$700 objects in this artificial population with $H_r<8$.

This is a very small population size.
We note that this specific analysis is only applicable to dynamically cold TNOs of relatively large size ($H_r<7$).
A steep size distribution could allow many smaller TNOs to remain undiscovered on similar orbits.
The point of this exercise it to show statistically tested constraints on populations with no survey detections.
For comparison, at this absolute magnitude limit ($H_r<7$), the scattering disk is estimated to have an intrinsic population of $\sim$4000 \citep{Lawleretal2018}, the plutinos are estimated to have an intrinsic population of $\sim$500 \citep{Volketal2016}, and the detached TNOs are estimated to number $\sim$4000 \citep{Petitetal2011}.
Using the estimated populations and size distribution slopes from \citet{Petitetal2011} gives 3000 $H_r<7$ TNOs in the classical belt.
The 3:1 mean-motion resonance, which is located at a similar semimajor axis ($a=62.6$~AU), is estimated to have an $H_r<7$ population of $\sim$200 \citep{Gladmanetal2012}.
The 3:1 TNOs, however, are much more easily detected in a survey due to their higher eccentricities as compared with our artificial 60~AU population, thus a given survey would be sensitive to a larger $H$-magnitude for the 3:1 population than the 60~AU cold test population.  
This statistically tested population limit essentially shows that not very many low eccentricity, distant TNOs can be hiding from the OSSOS survey ensemble, especially when compared with other TNO populations.

\section{Conclusion}

Using TNO discoveries from well-characterized surveys and only analysing the goodness of fit between models and TNO discoveries after forward-biasing the models gives a statistically powerful framework within which to validate dynamical models of Kuiper belt formation. 
Understanding the effects that various aspects of Neptune's migration have on the detailed structure of the Kuiper belt not only provides constraints on the formation of Neptune and Kuiper belt planetesimals, but also provides useful comparison to extrasolar planetesimal belts \citep{MatthewsKavelaars2016}.
There are, of course, many lingering mysteries about the structure of the Kuiper belt, some of which may be solved by detailed dynamical simulations in combination with new TNO discoveries in the near future.

One such mystery is explaining the very large inclinations in the scattering disk \citep{Shankmanetal2016,Lawleretal2018} and inside mean-motion resonances \citep{Gladmanetal2012}. 
Some possible dynamical mechanisms to raise inclinations include rogue planets/large mass TNOs \citep{GladmanChan2006,SilsbeeTremaine2017}, interactions with a distant massive planet \citep{Gomesetal2015,Lawleretal2017}, and diffusion from the Oort cloud \citep{Kaibetal2009,Brasseretal2012}.
These theories may also be related to the recent discovery that the mean plane of the distant Kuiper belt is warped \citep{VolkMalhotra2017}.

Explaining the population of high pericenter TNOs \citep{SheppardTrujillo2016,Shankmanetal2017} is also difficult with current Neptune migration models.  
Theories to emplace high pericenter TNOs include dynamical diffusion \citep{Bannisteretal2017a}, dropouts from mean-motion resonances during grainy Neptune migration \citep{KaibSheppard2016,Nesvornyetal2016}, dropouts from mean-motion resonances during Neptune's orbital circularization phase \citep{PikeLawler2017}, interactions with a distant giant planet \citep{Gomesetal2015,BatyginBrown2016,Lawleretal2017}, a stellar flyby \citep{MorbidelliLevison2004}, capture from a passing star \citep{Jilkovaetal2015}, and perturbations in the Solar birth cluster \citep{BrasserSchwamb2015}.

Here we have highlighted only a small number of the inconsistencies between models and real TNO orbital data.
The level of detail that must be included in Neptune migration simulations has increased dramatically with the release of the full OSSOS dataset, containing hundreds of TNOs with the most precise orbits ever measured.
The use of the Survey Simulator will be vital for testing future highly detailed dynamical emplacement simulations, and for solving the lingering mysteries in the observed structure of the Kuiper belt.

\acknowledgements

SML gratefully acknowledges support from the NRC-Canada Plaskett Fellowship.
MTB is supported by UK STFC grant ST/L000709/1.

The authors acknowledge the sacred nature of Maunakea, and appreciate the opportunity to observe from the mountain.
CFHT is operated by the National Research Council (NRC) of Canada, the Institute National des Sciences de l'Universe of the Centre National de la Recherche Scientifique (CNRS) of France, and the University of Hawaii, with OSSOS receiving additional access due to contributions from the Institute of Astronomy and Astrophysics, Academia Sinica, Taiwan.
Data are produced and hosted at the Canadian Astronomy Data Centre; processing and analysis were performed using computing and storage capacity provided by the Canadian Advanced Network For Astronomy Research (CANFAR).


 \newcommand{\noop}[1]{}

\end{document}